# A Comparison of Bacterial Colonies Count from Petri Dishes Utilizing Hough Transform and Traditional Manual Counting

Areesha Rehman [1], Zikria Saleem[2], Jarrar Amjad[3], Syed Rehan Shah[4], Kamran Siddique[5]

[1]Department of Pharmacy Practice, Bahauddin Zakariya University, Multan, 60000, Punjab, Pakistan
areeshaaimraan@gmail.com

[2]Department of Pharmacy Practice, Bahauddin Zakariya University, Multan, 60000, Punjab, Pakistan
xikria@gmail.com

[3]Department of Computer Science, Kansas State University, Manhattan, KS 66506, USA.
jarrar@ksu.edu

[4]Department of Computer Science, National University of Modern Languages, Islamabad 44000, Pakistan.
rehan.shah@numl.edu.pk

[5]Department of Computer Science and Engineering, University of Alaska Anchorage, Anchorage, United States
ksiddique@alaska.edu

Corresponding Author:

*Areesha Rehman*

*areeshaaimraan@gmail.com*

**Abstract:** Bacterial colony enumeration is an essential stage in microbiological research, allowing susceptibility to antibiotics assessment, monitoring of the environment, and clinical diagnostics. Traditional manual counting methods are costly and susceptible to human mistakes, prompting the creation of automated detection systems. This research compares the efficacy of the Hough Circle Transform method for automated colony detection to hand counting of E. coli, S. aureus, and P. aeruginosa colonies on 200 petri plates. These bacteria are among the most clinically relevant pathogens, with E. coli frequently causing urinary tract infections, S. aureus connected with skin and bloodstream infections, and P. aeruginosa a significant issue in hospital-acquired infections. When colonies were counted automatically without visual correction, the mean difference from manual counts was 59.7%, with overestimation and underestimation occurring in 29% and 45% of cases, whereas S. aureus and P. aeruginosa had higher error rates. The proposed methodology achieved an overall accuracy of 95% for E. coli, 90% for S. aureus, and 84% for P. aeruginosa, with associated recall values of 95%, 91%, and 86%. The F-measure remained continuously high, ranging between 0.85 and 0.95. Regarding efficiency, manual counting required an average of 70 seconds per plate, while automated counting without and with visual correction took 30 seconds. Despite issues with segmentation in high-density plates, automated approaches offer a potential approach to high-throughput bacterial enumeration by decreasing labor-intensive operations while retaining adequate accuracy. Future research should enhance colony algorithmic segmentation and picture preprocessing approaches to improve detection performance, especially on crowded petri plates. .

## 1. Introduction

The introduction Escherichia coli (E. coli) is a Gram-negative, facultatively anaerobic bacteria found in the digestive tracts of humans and animals. While most strains are innocuous and beneficial to gut health, certain pathogenic types cause foodborne illnesses, urinary tract infections, and severe gastrointestinal ailments [1]. Pseudomonas aeruginosa is a Gram-negative, pathogenic bacteria that causes significant nosocomial infections. It has multidrug resistance, caused by efflux pumps, β-lactamase synthesis, and decreased transparency of the membrane, making infections complicated to manage [1]. Staphylococcus aureus is a Gram-positive bacterium widely found on human skin and the nasopharynx. While it is frequently a harmless commensal, it is also a significant opportunistic pathogen, producing infections ranging from skin and soft tissue to potentially fatal endovascular and organ infections. It is a considerable source of hospital-acquired infections, especially in surgical incisions and indwelling medical equipment, where it may form biofilms and spread via the bloodstream [2].

Accurate quantification of bacterial colonies is fundamental in microbiological research, supporting applications in clinical diagnostics, antibiotic susceptibility testing, food safety, and therapeutic development. Manual colony counting remains the traditional method; however, it is time-consuming, susceptible to human error, and inconsistent, particularly when handling high bacterial loads. However, in manual counting, the researcher can only count 20-25% of visible colonies [3]. Moreover, there are various bacterial colony counters that are commercially available but inaccessible to smaller research labs, educational institutions, or developing countries [4]. Advances in image processing and artificial intelligence have led to the emergence of automated colony counting methods aimed at improving accuracy, efficiency, and scalability [30]. The automatic detection is cost-efficient, open source, and available to any individual and laboratory. However, these approaches utilize image processing techniques such as adaptive thresholding, contour detection, and deep learning models to differentiate bacterial colonies from background noise [3]. Therefore, automated counting of bacterial colonies has achieved almost 80% accuracy using MATLAB [5].

Despite significant advancements, challenges persist, including colony overlap, misdetections, and variability in colony morphology, necessitating continuous refinement of detection algorithms. Python-based image processing has enabled the development of automated colony counting systems that rely on open-source libraries such as OpenCV, NumPy, and TensorFlow for image preprocessing, object detection, and deep-learning-based colony enumeration. OpenCV supports adaptive thresholding for contour identification and using the Hough Circle Transform for colony recognition. Therefore, deep learning methods such as Faster R-CNN and YOLO enhance segmentation and categorization [6].

However, such modern modelling requires high performance GPU for training, beside our implemented Hough transform technique does not require any high performing GPUs, the detection can be done on low cost CPUs. Hough Circle Transform is indeed one of the most reliable algorithms for detecting perfectly spherical objects or almost spherical objects, which makes it ideally suited for bacterial colony counting, where colonies are predominantly circular, even in the light of advance development of deep learning segmentation models [7].

This study introduces an automated bacterial colony counting method validated against manual colony counts to enhance accuracy and efficiency. A dataset of 600 bacterial images (200 per species: Escherichia coli, Staphylococcus aureus, and Pseudomonas aeruginosa) utilized to evaluate detection performance [15]. The proposed technique offers a cost-effective, scalable alternative for laboratory automation, addressing limitations in conventional colony counting methodologies.

The following are the research objectives:
- Evaluate the accuracy and efficiency of the Hough Circle Transform for bacterial colony counting.
- Compare automated counting results with manual colony enumeration.
- Analyze the impact of colony density and image quality on counting accuracy.

Assess potential improvements in automated detection through preprocessing enhancements.

A low-cost, high-throughput automated colony counting device that used consumer-grade digital cameras or document scanners with specialized software. This method drastically decreased expenses and improved accessibility for laboratories looking for fast colony counting options. While this technique increased performance, issues like reliably recognizing overlapping colonies and separating colonies from background artifacts persisted [8]

Automated colony counting using MATLAB and image processing achieved an overall accuracy of 80%. The study did not incorporate the Hough Transform method with manual counting methods, highlighting the key advantages over the traditional or labor-intensive techniques [11]

There are numerous ways to collect bacteria colonies, including neural networks using convolution (CNNs) on microbiology plates. The researchers developed a technique using CNN to automate the counting process to decrease the laborious and susceptible to errors nature of hand counting. This method uses image processing methods in conjunction with deep learning algorithms to properly locate and count bacterial colonies, emphasizing the CNN algorithm's potential for increasing the effectiveness and precision of microbiological examinations [6].

The detection of bacterial colonies using image processing decreases labor costs and increases efficiency. The article's objective is to present the automation of a method for counting colonies of bacteria using YOLOv5 TensorRT on Jetson Nano instead of Hough Transform. It compared human counting, as the manual technique, and automated tools, where it sets a mean error value of 0.0142%, demonstrating accuracy [12].

The article by [6] demonstrated an overall F1 score of 86% by detecting bacterial colonies using the YOLOv4 model. The authors proposed three more deep learning methods showing the potential use of automatic methods in medical sciences. Their focus was on detection and lacked the Counting and use of Hough Transform.

A computer vision method was incorporated to count microbial colonies, showing less than 10% deviation in Baird Parker plates when it comes to manual counting; however, the researcher was unable to visualize greater discrepancies observed from PCA plates. However, both methods worked efficiently and showed no statistically significant difference measured [13].

According to the research by [5], the use of deep learning in microbiology counting lays an efficient ground for testing bacterial cells and contributes to antibody research. The authors used high-resolution images to segment the colony count, which showed an 87% score in bright field counting.

The paper by [7] proposed a semi-automatic image processing system to overcome manual counting. The difference shows the mean accuracy of 82%, 86%, and 85%, respectively. Their automatic low-cost system provided an enumeration process where they extracted image features and tested them with more than 300 colonies.

## 2. Materials and Methods
### 2.1. Sample Collection and Image Preparation

The bacterial colonies (Escherichia coli, Staphylococcus aureus, and Pseudomonas aeruginosa) were grown on agar plates under typical microbiological laboratory conditions. According to [7] all the agar plates were incubated at around 37°C for about 24 hours to help grow colonies. A digital camera is used to obtain high-resolution pictures of the agar plates in a jpeg format under controlled lighting environments, ensuring consistent lighting and reducing glare.

### 2.1.2. Image Dataset Preparation

In this study, 600 pictures were obtained (200 of each bacterial species). Each image have been manually categorized into folders based on the bacteria species Escherichia coli, Pseudomonas aeruginosa, and Staphylococcus aureus [18]. All photos were scaled to 512 × 512 pixels in the preprocessing stage with Python's PIL module. The Hough Circle was the feature detection technique employed, as shown in Figure 1.

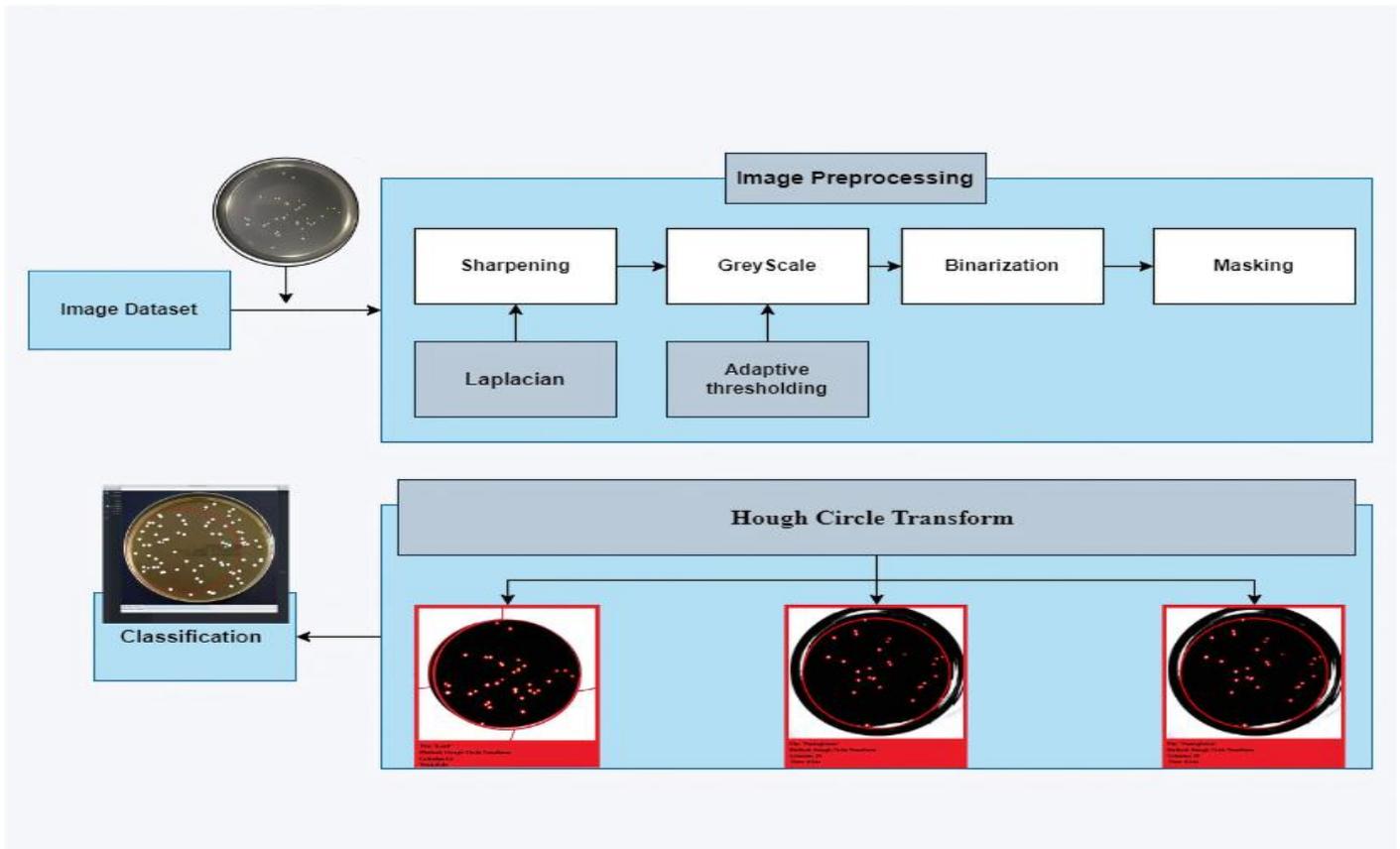

**Figure 1**: Overall flow of methodology.

**2.2. Bacterial Colony Annotation**

An annotation procedure is adopted to identify bacterial colonies precisely in pictures of Petri dishes. Following are the annotation steps:

Manual Annotation: One of the most common responsibilities in microbiology labs is to count the bacterial colonies growing on agar plates. Manual counting is often accomplished using a writing instrument to mark as a bound box and an electronic clicker device to maintain the correct path. This procedure is inexpensive and straightforward, but it is also susceptible to human errors, as according to the plate dilution, there might be up to 1000 colonies present. Manual annotation burdens the microbiologist during the counting procedure. This counting procedure appears to be a time-consuming and unpleasant operation [32].

Automated Annotation: The enhanced way to find circular colonies in the photos was the Hough Circle Transform technique, which helped extract bacterial colonies and coordinate detected colony edges [19]. For compatibility with different object identification frameworks, the automatically discovered bounding boxes formatted in YOLO format before being changed to COCO format Figure 2.

**2.3. Dataset Formatting and Conversion**

The number of manually counted colonies identifies as the ground truth for assessing the automated Hough Circle Transform approach to maintain analytical consistency. The dataset was created by compiling annotations from both manual and automatic approaches in a structured manner. Manual counting includes details of picture filenames, image size, bacterial genus, and discovered colony positions. The prepared data was then converted to Python-based OpenCV analysis tools for additional validation and statistical analysis [20].

**2.4. Image Preprocessing**

Before executing the Hough Circle Transform, we implemented image preprocessing to increase the accuracy of colony detection:

Preprocessing Steps: To improve the contrast between bacterial colonies and the agar plate background, the raw pictures were processed using adaptive thresholding, Laplacian sharpening, and grayscale conversion, which can be visualized in Figure 2.

Application of Hough Circle Transform: OpenCV's Hough Circle Transform was used to evaluate the preprocessed images and identify circular shapes that were suggestive of bacterial colonies.

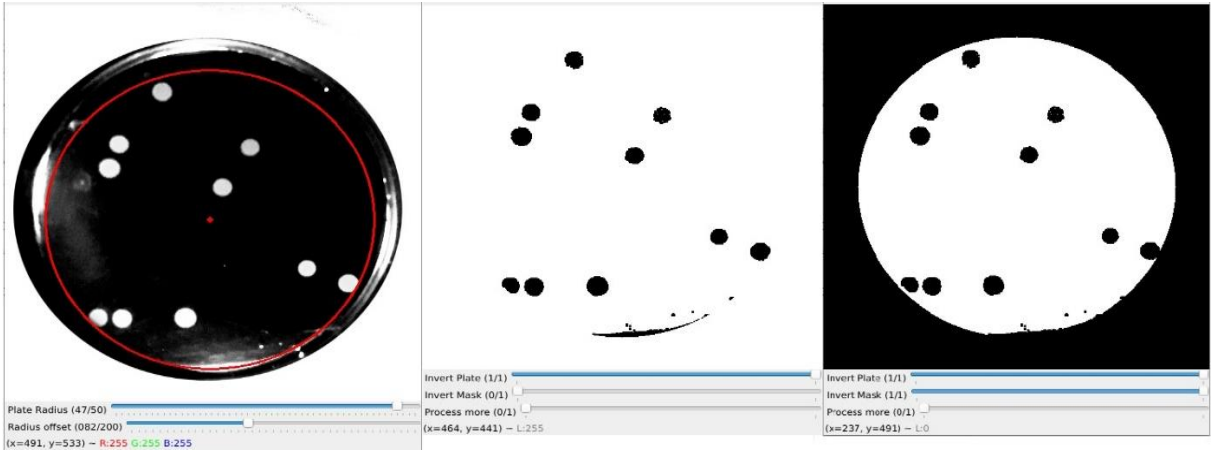

**Figure 2**. Preprocessed bacterial colony.

### 2.5. Hardware Setup

Preferred Hardware and software will be used for all tests listed below:

- OS: Windows 11
- Processor: Intel ® Core ™ i5-8145U CPU @ 2.10 GHz
- Ram: 16 GB
- IDE: Visual Studio Code
- Language: Python 3.12

### 2.6. Analysis of Statistics

Statistical comparisons made to evaluate the artificial colony detection method's efficacy:

- Analysis of Colony Counts: A statistical comparison of automatic Hough Circle Transform detections and manual colony counts conducted. Error rate graphs and histograms used to display the findings.
- Correlation Analysis: The relationship between manually counted colonies and automated detections was measured using Pearson's correlation coefficient. A high correlation would demonstrate the computerized approach's dependability.

**Evaluation and Performance Metrics**

The following crucial metrics used to evaluate the Hough Circle Transform method's performance:

- Accuracy: The percentage of correctly identified colonies compared to all colonies present is known as accuracy [28].

    Accuracy=True Positives+False Positives+False Negatives/True Positives     (1)

- Recall: The proportion of actual colonies that the automated approach was able to identify.

    Recall=True Positives+False Negatives/True Positives     (2)

- F1-score: The F1-score is a balanced indicator of colony detection performance calculated as the mean of both recall and accuracy.

    F1-score=2×Precision+Recall/Precision×Recall     (3)

- Error Rate: The error rate calculated using $C(x)C(x)C(x)$ is the number of colonies detected by the technique at a specific plate density. $FPR(x)FPR(x)FPR(x)$ is the False Positive Rate. $FNR(x)FNR(x)FNR(x)$ is the False Negative Rate.

$$\text{Error}(x) = C(x) \times (FPR(x) + FNR(x)) \qquad (4)$$

## 3. Results
### 3.1. Automated Detection Representation
The graphic showed the detailed process of detecting bacterial colonies on Petri plates using the Hough Circle Transform technique. The original petri dish photos Figure 3 A show bacterial colonies of E. coli, S. aureus, and P. aeruginosa. The colonies are segmented, binarized, and then discovered using automatic contouring Figure 3 B. Final processed photos represent recognized bacterial colonies, indicating the algorithm's ability to recognize various colony shapes Figure 3 C.

### 3.2 Implication of image processing pathway
- Preprocessing: The incoming picture is converted to grayscale and thresholded to improve contrast.
- Edge Detection and Separation: Adaptive filtering and binary conversion separate bacterial colonies while reducing background noise [27].
- Hough Circle Transformation: This approach recognizes circular characteristics that correlate to bacterial colonies and labels them with red outlines.
- Quantification and Accuracy Check: The overall colony count is presented after comparing the identified colonies to the original petri dish picture.

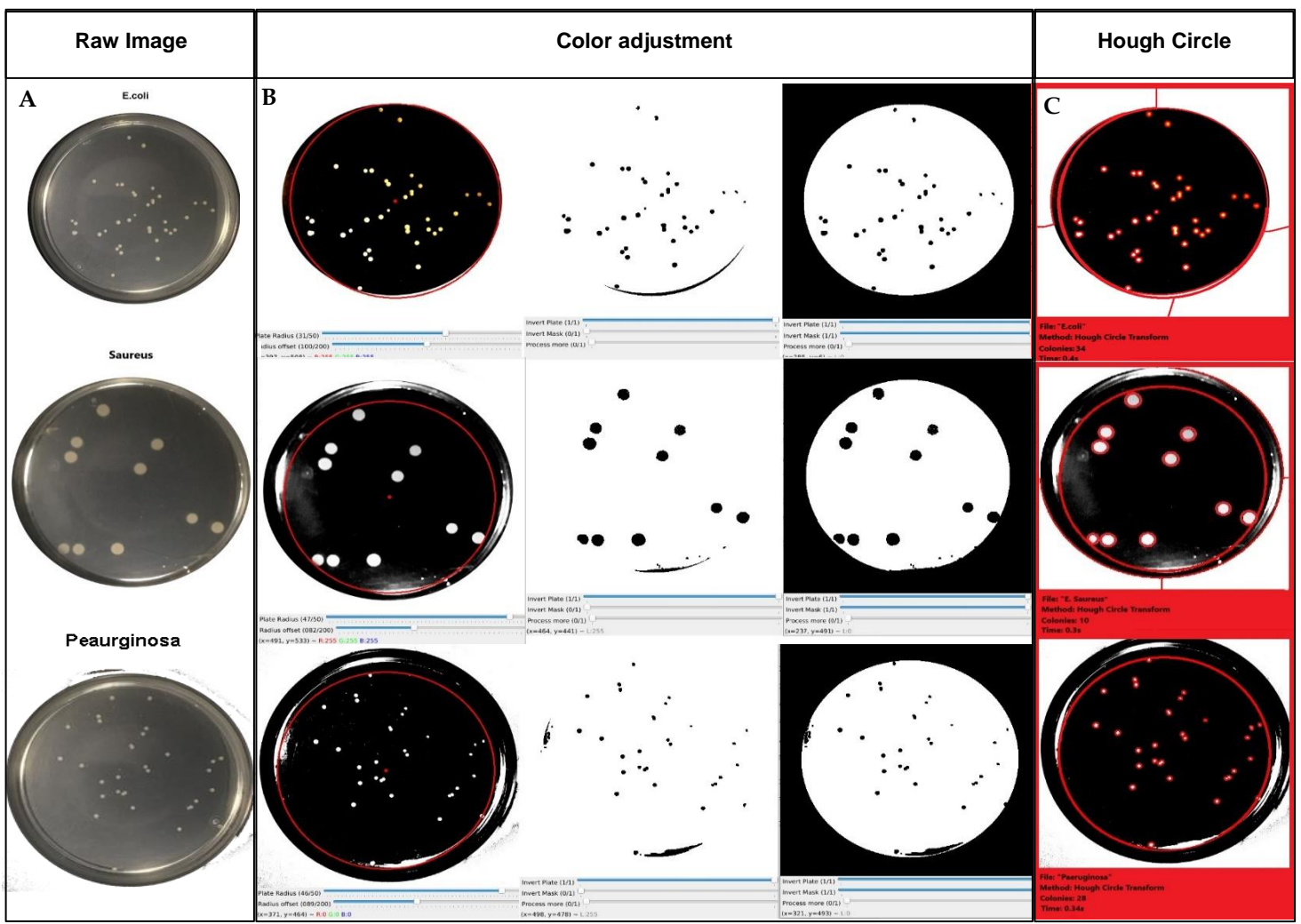

**Figure 3.** Application of Hough circle transform after original raw image adjusted into grayscale image and preprocessing (Rodrigues et al., 2022).

## 3.3. Hough Circle Transform

The Hough Circle Transform's performance in automated bacterial colony counting was assessed by comparing its results to manual counting across three bacterial species: E. coli, S. aureus, and P. aeruginosa Figure 2. The error rate observation shows that automatic detection nearly matches hand counting at lower colony densities, with slight differences. The Equation 5 showing (a,b) is the coordinate of the Center of the , (r) is the radius of the circle and (x,y) stand for any point on the circle.

$$R^2 =(x-a)^2+(y-b)^2 \qquad (5)$$

However, as bacterial populations grow, particularly above 100 colonies per plate, the error rate rises due to colony overlap and segmentation problems, resulting in under- or over-counting. Among the three bacteria, E. coli had the least error variation, implying that its colony morphology is more distinct by the algorithm than S. aureus and P. aeruginosa, where clustering and irregular colony formations affect detection Accuracy. In Hough transform, to find circles, points in the image are mapped to some parameter space for each point (x,y) to vote for all possible circle canters and (a,b) as a function of radius r. The Equation 6 and 7 showing the angle theta ranges between 0 and 360, this meaning that it defines all the points that constitute a perfect circle. Each endpoint in the image casts votes for any conceivable canter (a,b) of a circle in a 3-d parameter space: (a,b,r).

$$a=x-r\cos(\theta) \qquad (6)$$
$$b=y-r\sin(\theta) \qquad (7)$$

Furthermore, picture quality issues such as inadequate lighting and blurry sections caused misdetection in high-density plates. When colonies were counted automatically without visual correction, the mean difference from manual counts was 59.7%, with overestimation and underestimation occurring in 29% and 45% of cases, whereas S. aureus and P. aeruginosa had higher error rates in 300 colonies [8]. Still, the Hough Circle Transform accomplished the operation in less than a minute, despite the colony count (Fig 3). This consistency demonstrates the possibility of automated detection for high quantities of bacterial Script, which can reduce labor-intensive human counting while remaining efficient. However, more improvement in colony identification and separation approaches is required to improve accuracy, especially on high-density plates [26].

## 4. Discussion

The findings of this study are consistent with previous research works in the area of automated bacterial colony count, particularly those that use the Hough Circle Transform for colony detection, as the research worker suggested an agar plate solution for bacterial colony counting. The study observed that identification error are more pronounced when bacterial density in Petri dishes grew Figure 4.

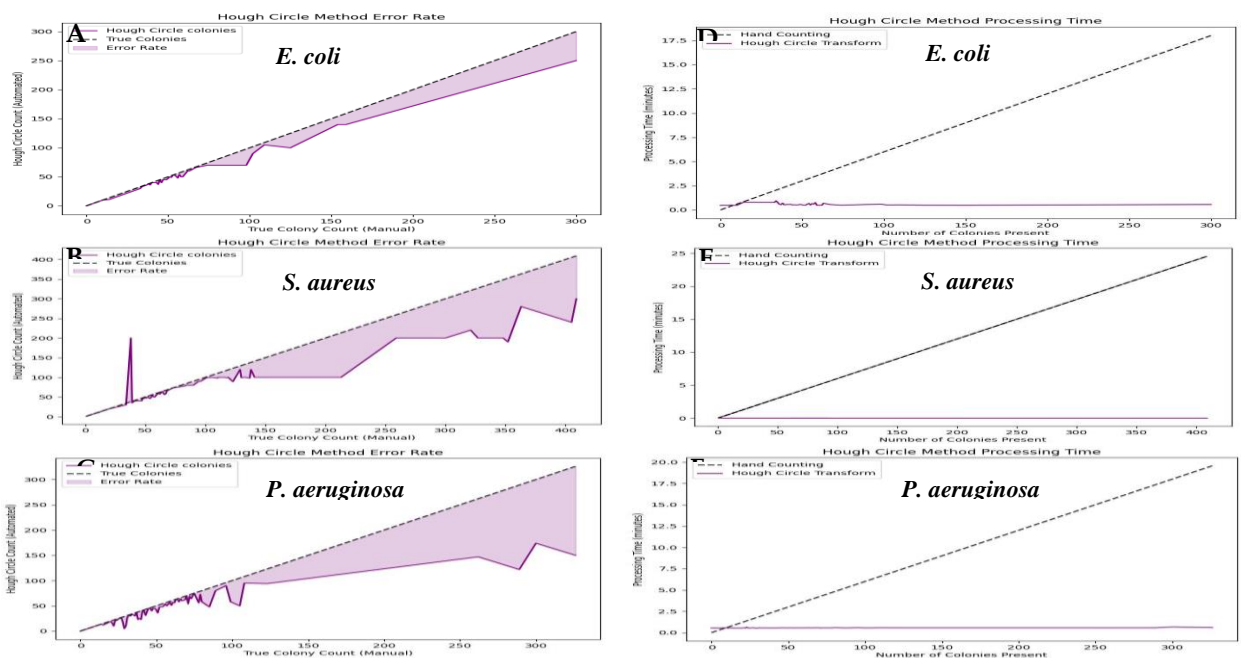

**Figure 4.** Line graph of the Hough Circle Transform on E. coli, S. aureus, and P. aeruginosa bacteria with comparison of manual and automated colony counting for with error rate analysis indicates minor deviation at low colony counts (A-C) and comparison of processing time (D-F) showed automation's efficiency

This supports prior studies that found segmentation issues when colonies cluster tightly together [25]. Furthermore, processing time analysis improves the efficiency of automated detection since the Hough Circle Transform maintains a consistent calculation time, significantly lowering microbiologists' human burden [14]. These findings support the potential of open-source automated colony counting systems; nevertheless, additional refining, notably in preprocessing approaches to reduce picture noise and increase colony separation, is required for widespread laboratory deployment [6].

The research found that as the number of bacteria in the Petri dish medium increases, separation errors become more evident due to colony overlap, leading to missing and overestimating. This conclusion corresponds with earlier research showing that grouping significantly impacts the precision of detection. Furthermore, the value for the Hough Circle Transform demonstrated maintaining consistent time for computation, resulting in a substantial reduction in manual workload [14]. While the process is effective, it requires more refinement to maximize colony separation, particularly for a high level of populations of bacteria [24].
Future developments in preprocessing methods and image noise mitigation may enhance identification accuracy, rendering this technique more appropriate for widespread laboratory application [6].

**4.1. Histogram of Bacterial colony counts in Hough Circle Transform Method**

The histograms observed the frequency range of bacterial colony counts for E. coli, S. aureus, and P. aeruginosa with the automated Hough Circle Transform method Figure 5 A-B-C.

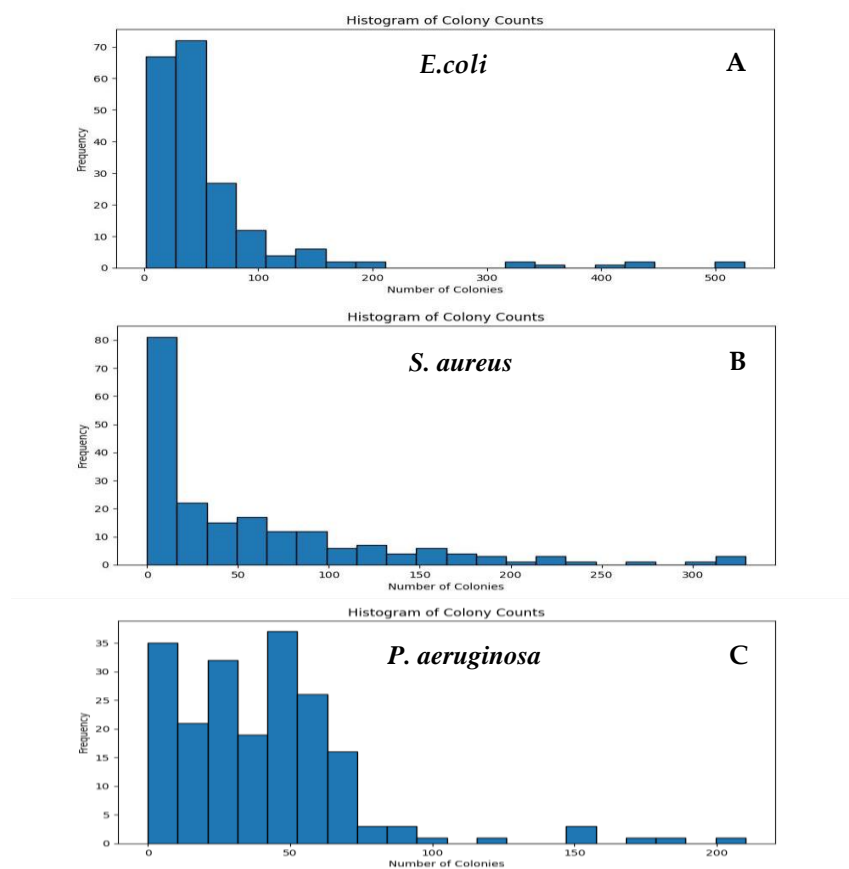

**Figure 5.** Histogram of bacterial colony counts on Petri plates using the Hough Circle Transform approach for E. coli (A), S. aureus (B), and P. aeruginosa (C). The distribution pattern demonstrated the frequency of observed colonies and changes in colony density and recognition precision across bacterial species.

The distribution trends show that lower colony numbers are more common, whilst larger colony counts are more variable [14]. The automated technique showed greater accuracy at lower colony densities but had difficulty with higher bacterial populations. The increased variability in colony counts at greater densities suggests difficulties in automated detection due to overlapping colonies, segmentation issues, and potential misclassification. Differences in distribution density across species of bacteria showed modifications to colony structure and the clustering patterns, which may influence the precision of detection.

The presence of irregular distributions emphasizes the importance of enhancing automated segmentation techniques for high-density colony conditions. These outcomes emphasize the demand for better image preparatory methods, including more significant brightness adjustment and clearing, to eliminate errors resulting from low-resolution images. Our studies, which were related to previous studies, demonstrated the distribution of bacterial colony counts detected using the Hough Circle Transform toward hand counting. A substantial proportion of frequencies in a particular interval throughout the distribution of values implies greater detection consistency, meaning that the method accurately detects a similar amount of colony across several pictures. This finding is identical to a recent investigation, which discovered that automated algorithms for segmentation perform well while minimal colony overlaps [23].

## 4.2. Pearson Correlation

Pearson's correlation coefficient analyzes the relationship between manual and automated detection of bacterial species: E. coli, S. aureus, and P. aeruginosa , as shown in Figure 6.

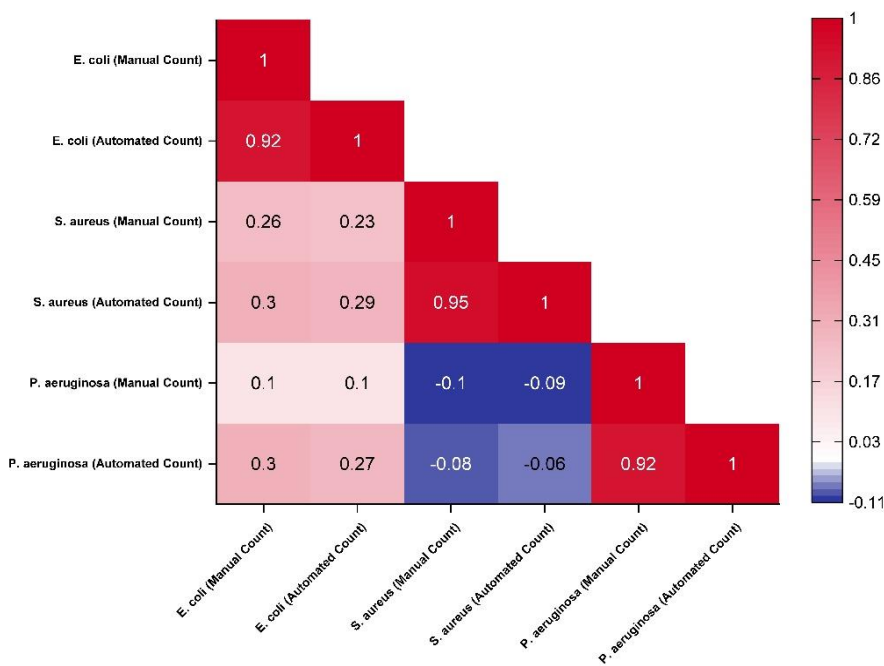

**Figure 6**. Pearson's correlation heatmap compared manual and automated colony counts for E. coli, S. aureus, and P. aeruginosa. Higher scores indicate a better agreement between manual and automated identification techniques.

The correlation study revealed that the computerized counting technique was more reliable and consistent than regular manual counting. The correlation heatmap revealed a significant positive correlation between traditional and programmed E. coli and P. aeruginosa counts, indicating good accuracy in automated detection.
Similarly, S. aureus demonstrated a moderate-to-strong correlation, indicating that the computerized technique resembles some variability in manual counting. Lower correlation scores in specific circumstances indicate potential limitations of the programmed method that might exist and be attributed to overlapped colonies, processing of image issues, or variances in the structure of colonies. Overall, these data lend support to the automated detection method for bacterial colony enumeration, demonstrating its ability to perform a speedy and reliable microbiological examination while also highlighting possibilities for additional improvement as a recent paper also correlates the data with the automated method by MATLAB [7].

**4.3. Assessment of the Hough Circle Transform method as an automated Colony Counting compared with the Traditional counting Method**

The performance of the Hough Circle Transform-based automated colony counting method was compared to manual counting across E. coli, S. aureus, and P. aeruginosa colony counts, as indicated in Table 1. In the automatically counted colonies without visual correction, the mean difference from manual counts was 59.7% [8], with overestimation and underestimation occurring in 29% and 45% of cases, whereas S. aureus and P. aeruginosa had higher error rates. The proposed methodology achieved an overall accuracy of 95% for E. coli, 90% for S. aureus, and 84% for P. aeruginosa, with associated recall values of 95%, 91%, and 86%. The F-measure values for the automated method were 0.93, 0.88, and 0.81, demonstrating good agreement with manual counting but exposing certain limits in densely filled petri plates. Automated counting has lower accuracy, notably for E. coli and S. aureus, due to overlapping colonies [22], changes in bacterial morphology, and picture quality concerns such as blurring and uneven lighting conditions. These variables influence the segmentation process, resulting in underestimating or overestimating colony numbers. However, the greater recall values indicate that the automated technique is excellent at recognizing the majority of colonies but may have difficulty with the exact difference in enclosed petri plates. Although manual counting remained the preferred method, the automated technique performed well and provided significant benefits concerning efficiency and scalability, especially for high-throughput microbiological tests. Further enhancements to image processing algorithms and separation approaches may increase automated detection precision, especially in high-density bacterial cultures [10].

The effectiveness of automated techniques, especially the Hough Circle Transform, in colony enumeration is demonstrated by comparing accuracy, recall, and F-measure between automated and manual detection. Our study related to the recent research that automated image analysis techniques save time and effort compared to manual counting since they offer high accuracy with little error. For E. coli, which had the lowest mistake rate among the studied species, the automated method in this study proved to be as accurate as hand counting. However, because of colony clustering and overlapping problems, P. aeruginosa still showed poorer precision even if the automated technique showed higher accuracy. Although the computerized technique may sometimes under-segment colonies in high-density locations, the recall values indicate that it is effective at finding colonies [21].

**Table 1**. Indicates the accuracy, recall, and F-measure values of bacterial colony counts using traditional manual and automated methods. The F-measure provides an accurate evaluation of precision and recall, while correctly identified colonies showed accuracy and denoted with percentages.

| Bacteria | Accuracy (0-200) Automated | Recall (0-200) Automated | F-measure (0-200) Automated | Accuracy (0-200) Manual | Recall (0-200) Manual | F-measure (0-200) Manual |
|---|---|---|---|---|---|---|
| E. coli | 95% | 95% | 0.95 | 85% | 90% | 0.93 |
| S. aureus | 90% | 91% | 0.90 | 84% | 89% | 0.88 |
| P. aeruginosa | 84% | 86% | 0.85 | 90% | 93% | 0.81 |

**5. Conclusions**

The outcomes of this research showed that the Hough Circle Transform approach is more effective for automated bacterial colony detection than manual counting. The mean difference from manual counts was 59.7%, with overestimation and underestimation occurring in 29% and 45% of cases, whereas S. aureus and P. aeruginosa had higher error rates. The proposed methodology achieved an overall accuracy of 95% for E. coli, 90% for S. aureus, and 84% for P. aeruginosa, with associated recall values of 95%, 91%, and 86%. The F-measure remained continuously high, ranging between 0.85 and 0.95. thereafter P. aeruginosa and S. aureus, with no significant variations in recall and F-measure values. Automated detection retains excellent accuracy while showing minor decreases in recall and F-measure, especially in high-density colony plates. The data additionally illustrated that hand counting is still extraordinarily accurate but requires a long time, whereas automated methods facilitate faster processing and constant detection rates. The test data set of 600 photos of petri plates by cell phone had slight variance, resulting in slightly contrasted agar color and bacteria changes.

The images were low quality and susceptible to distortion via glare and blur. Hand-counted colony methods are used as the primary source of accuracy for optimal accuracy in test observations. While this is acceptable compared to methods, it can lead to inaccurate scaling when comparing results to previous studies. Hand counting is susceptible to human error. Future enhancements to the Watershed technique might include automating plate recognition, which takes up most of the time. This innovation might make the watershed approach very instantaneous. Testing on a bigger dataset with varying pigmentation and opacity levels for agar and bacteria might improve preprocessing approaches and create "presets" for varied contrast levels. To improve user-friendliness, consider optimizing OpenCV's debug windows with graphical libraries like Tkinter or developing a mobile app for smartphone testing.

**Data Availability Statement:** The data is publically available at:  https://doi.org/10.3390/bioengineering9070271

**Conflicts of Interest:** The authors declare no conflicts of interest.